\documentclass[preprint,onecolumn,nobibnotes,nofootinbib,showpacs,preprintnumbers,superscriptaddress,amsmath,amssymb]{revtex4}
\pdfoutput=1
\usepackage{booktabs,graphicx,pstricks,verbatim,url}
\newcommand{\be}{\begin{eqnarray}}
\newcommand{\ee}{\end{eqnarray}}
\def\ltap{\ \raise.3ex\hbox{$<$\kern-.75em\lower1ex\hbox{$\sim$}}\ }
\def\gtap{\ \raise.3ex\hbox{$>$\kern-.75em\lower1ex\hbox{$\sim$}}\ }
\def\lsim{\ \raise.3ex\hbox{$<$\kern-.75em\lower1ex\hbox{$\sim$}}\ }
\def\gsim{\ \raise.3ex\hbox{$>$\kern-.75em\lower1ex\hbox{$\sim$}}\ }

\newcommand{\kev}{{\rm keV}}

\newcommand{\gev}{{\rm GeV}}

\newcommand{\kevee}{{\rm keVee}}
\begin{document}
\begin{flushright}
FERMILAB-PUB-10-437-T
\end{flushright}

\title{Integrating Out Astrophysical Uncertainties}
\author{Patrick J. Fox}
\affiliation{Theoretical Physics Department, Fermilab, Batavia, Illinois 60510, USA}
\affiliation{School of Natural Sciences, Institute for Advanced Study, Einstein Drive, Princeton, NJ 08540}
\author{Jia Liu}
\affiliation{Center for Cosmology and Particle Physics, Department of Physics, New York University, New York, NY 10003, USA}
\author{Neal Weiner}
\affiliation{Center for Cosmology and Particle Physics,  Department of Physics, New York University, New York, NY 10003, USA}
\affiliation{School of Natural Sciences, Institute for Advanced Study, Einstein Drive, Princeton, NJ 08540}
\begin{abstract}
Underground searches for dark matter involve a complicated interplay of particle physics, nuclear physics, atomic physics and astrophysics. 
We attempt to remove the uncertainties associated with astrophysics by developing the means to map the observed signal in one experiment directly into a predicted rate at another. We argue that 
it is possible to make experimental comparisons that are completely free of astrophysical uncertainties
by focusing on {\em integral} quantities, such as $g(v_{min})=\int_{v_{min}} dv\, f(v)/v $ and $\int_{v_{thresh}} dv\, v g(v)$. Direct comparisons are possible when the $v_{min}$ space probed by different experiments overlap. As examples, we consider the possible dark matter signals at CoGeNT, DAMA and CRESST-Oxygen. We find that expected rate from CoGeNT in the XENON10 experiment is higher than observed, unless scintillation light output is low. Moreover, we determine that S2-only analyses are constraining, unless the charge yield $Q_y< 2.4 {\, \rm electrons/keV}$. For DAMA to be consistent with XENON10, we find for $q_{Na}=0.3$ that the modulation rate must be extremely high ($\gsim 70\%$ for $m_\chi = 7\, \gev$), while for higher quenching factors, it makes an explicit prediction (0.8 - 0.9 cpd/kg) for the modulation to be observed at CoGeNT. Finally, we find CDMS-Si, even with a 10 keV threshold, as well as XENON10, even with low scintillation, would have seen significant rates if the excess events at CRESST arise from elastic WIMP scattering, making it very unlikely to be the explanation of this anomaly.
\end{abstract}
\maketitle

\section{introduction}
The search for dark matter is a central priority of modern high energy physics. Understanding the nature of the $\sim$24\% of the universe composed of dark matter would give insight into the origin of galaxies and cosmic structures, the universe at high temperatures and a broader particle physics context for the Standard Model. To this end, a wide range of underground detectors have been and will be deployed to search for it (e.g., \cite{Behnke:2008zza,Angloher:2008jj,Lebedenko:2008gb,Ahmed:2009zw,Sanglard:2009qp,Bernabei:2010mq,Aalseth:2010vx,Aprile:2010um,Piro:2010qd}). These programs utilize many different technologies and targets, and have sensitivities to a number of different WIMP scenarios, as well as mass ranges and interaction properties.

These rare-event searches are some of the most sensitive detectors ever built, seeking a signal that may be as small as a few events per year. Consequently, they are sensitive to new and unexpected backgrounds as well.  Any claim of dark matter discovery must be confirmed with multiple technologies before it can be believed. At the same time, comparing different experiments is a great challenge, with uncertainties from particle physics, nuclear physics, atomic physics and astrophysics compounding one another.

Particle physics uncertainties can be explored by considering phenomenologically varied models, such as spin-dependent \cite{Jungman:1995df} or momentum-dependent \cite{Feldstein:2009tr,Chang:2009yt} couplings, inelastic \cite{TuckerSmith:2001hy,TuckerSmith:2004jv} scattering, electromagnetic charge radius or dipole interactions \cite{Bagnasco:1993st,Pospelov:2000bq,Sigurdson:2004zp,Barger:2010gv,Chang:2010en,Fitzpatrick:2010br,Banks:2010eh}, resonant dark matter \cite{Bai:2009cd}, mirror matter \cite{Foot:2003iv,Foot:2010hu} among others. Atomic and nuclear physics uncertainties can be better understood through theory and careful experiment (e.g., \cite{Duda:2006uk,Sorensen:2008ec,Manzur:2009hp}).  Astrophysical uncertainties are a challenge, however, for a number of reasons.

High resolution numerical studies \cite{Vogelsberger:2008qb,Kuhlen:2009vh,Ling:2009eh} have confirmed that Maxwellian distributions are generally good approximations to the phase space structure of dark matter halos. Nonetheless, significant deviations are found at high velocities that can be relevant for some scenarios, such as light WIMPs or inelastic WIMPs \cite{MarchRussell:2008dy,Kuhlen:2009vh,McCabe:2010zh}, and in analytic solutions to NFW profiles such deviations are calculable \cite{Lisanti:2010qx}. Although the resolution of the numerical studies is not adequate to study small-scales accurately, variations from place to place within a halo can be important \cite{Kuhlen:2009vh}. Pronounced structures such as streams or subhalos can dramatically alter expectations \cite{Freese:2003tt,Savage:2006qr,Lang:2010cd,Alves:2010pt}. And of course, the ``unknown unknowns'' are impossible to quantify.

It is important therefore to find methods that constrain scenarios without appealing to {\em any} model of the dark matter distribution. Some efforts at this have been studied already. For instance, \cite{Chang:2008xa} argued that an independent comparison for the iodine spin-independent explanation of DAMA could be made by studying the comparable range of energy at a Xenon target, given their kinematical similarity. It was pointed out in \cite{Chang:2010yk} that there is an overlap in velocity space between the $\sim 1$keVee signal at CoGeNT and the 7 keVr threshold at CDMS-Si. With positive results at two experiments, a measurement of the WIMP mass can be done without assuming a halo model \cite{Drees:2008bv}. Finally, \cite{Fox:2010bu} studied the possibility of extracting $f(v)$ from dark matter experiments in the future when large signals have been found.

In this paper, we take a different approach. Rather than attempt to find the physical function $f(v)$, or study variations in it, we attempt to directly map experimental signals from one detector to another. We do this by focusing on {\em integral} quantities, namely $g(v_{min}) = \int_{v_{min}}dv f(v)/v$ and $\int  dv\, v g(v)$. We determine the robustness of constraints by considering the relationship between recoil energy and $v_{min}$ space, rather than actual velocity space.  Although in our approaches we will gain less information about astrophysics, we can compare experiments even when $f(v)$ cannot be reliably extracted.

\section{$v_{min}$ ranges and astrophysics-independent scattering rates}
\label{sec:formalism}
Our approach will be simple: we will endeavor to map an energy range in a given experiment into the halo velocity space, and from there into any other experiment we wish to compare to. In this way, we can determine what energy ranges of experiments can be directly compared. In optimal situations, we will be able to extract $g(v)$, while in less optimal situations we will only be able to discuss total rates.

We begin with the differential rate at a direct detection experiment, which for elastically scattering DM is given by,
\be
\frac{dR}{dE_R} = \frac{N_T M_T \rho}{2 m_\chi\mu^2}  \sigma(E_R)\, g(v_{min})~,
\label{eq:rate}
\ee
where $\mu$ is the DM-nucleus reduced mass, and $N_T=\kappa N_A m_p/M_T$ is the number of target scattering sites per kg with $N_A$ Avogadro's number and $\kappa$ the mass fraction of the detector that is scattering DM.  The function $g(v_{min})$ is related to the integral of the DM speed distribution\footnote{It is usually assumed that the DM follows a Maxwell-Boltzmann distribution (in the galactic frame), with characteristic speed $v_0$, in which case (again in the galactic frame) $f(v)\propto v^2 e^{-v^2/v_0^2}$.}, $f(v,t)$, by, 
\be
g(v_{min},t)=\int_{v_{min}}^\infty dv\,\frac{f(v,t)}{v}~.
\label{eq:defgofv}
\ee
There is a minimum speed that the DM must have in order to deposit recoil energy $E_R$ in the detector.  For elastically scattering WIMPs this minimum velocity is
\be
v_{min} = \sqrt{\frac{M_T E_R}{2 \mu^2}}~.
\label{eq:vmin}
\ee

Making a comparison between different experiments is confused by the fact that it is not a single velocity that contributes to the scattering rate at a particular $E_R$. Rather, {\em all} particles with velocities greater than $v_{min}$ will contribute, making it impossible to map rates into velocity space.

However, we can consider a related space -- $v_{min}$-space, whose elements are the sets of all particles with velocities greater than $v_{min}$. Because all particles with adequately high velocities contribute, it {\em is} reasonable to consider a mapping between $E_R$ and $v_{min}$ through  (\ref{eq:vmin}). 

This simple relationship allows us to compare results from different direct detection experiments without making an assumption about the distribution of DM velocities in the Milky Way's halo, provided one can relate the scattering cross sections at the various experiments.  In the standard cases of SI or SD DM the nuclear scattering cross section can be related to the nucleonic (in this case the proton) cross section as
\be
\sigma_{SI}(E_R)&=&\sigma_p \frac{\mu^2}{\mu_{n\chi}^2}\frac{\left(f_p\, Z+f_n\, (A-Z)\right)^2}{f_p^2} F^2(E_R) \label{eq:xsecSI}\\
\sigma_{SD}(E_R)&=& \frac{\sigma_p}{2J+1} \frac{\mu^2}{\mu_{n\chi}^2} \frac{\left(a_p^2\, S_{pp}(E_R)+a_p\,a_nS_{pn}(E_R)+a_n^2 S_{nn}(E_R)\right)^2}{a_p^2}  ~,
\label{eq:xsecSD}
\ee
allowing comparison of different experiments, we have defined $\mu_{n\chi}$ as the DM-nucleon reduced mass.  

Let us suppose we have two experiments to compare, with targets $T_{1,2}$ with masses $M_{1,2}$. We assume the first has a signal which appears over an energy range $[E^{(1)}_{low},E^{(1)}_{high}]$.  This energy range correspond to $v_{min}$ ranges $[v^{low}_{min},v_{min}^{high}]$, using (\ref{eq:vmin}).  

This brings us to the central point of our efforts: to make a comparison between two experiments one must first determine whether the {\em $v_{min}$} space probed by the two experiments overlaps. As a matter of practical course, a given experiment has a lower energy threshold $E_{min}$, which can be translated into a lower bound on the $v_{min}$ range.  If experiment 1 has data for the differential rate of DM scattering in their experiment, $dR_1/dE_R$ at energies $E_i^{(1)}$ this can be used to predict a rate at energy $E_i^{(2)}$ at experiment 2, $dR_2/dE_R$, or \emph{vice versa} if experiment 2 has the signal.  Thus, we have
\be
[E_{low}^{(1)},E_{high}^{(1)}] \Longleftrightarrow [v^{low}_{min},v^{high}_{min}] \Longleftrightarrow [E_{low}^{(2)},E_{high}^{(2)}],
\label{eq:eminrangemap}
\ee
where 
\be
[E_{low}^{(2)},E_{high}^{(2)}] = \frac{\mu_2^2 M_T^{(1)}}{\mu_1^2 M_T^{(2)}}[E_{low}^{(1)},E_{high}^{(1)}].
\label{eq:emap}
\ee

We can invert (\ref{eq:rate}) to solve for $g(v_{min})$ \emph{limited to the range} 
$v_{min} \in [v^{low}_{min,1},v^{high}_{min,1}]$
\be
g(v_{min}) = \frac{2 m_\chi\mu^2} {N_A\kappa\, m_p\,\rho\,  \sigma(E_R)}\frac{dR_1}{dE_1}
\label{eq:geq}
\ee
This then allows us to explicitly state the expected rate for experiment two, again\footnote{Since $g(v)$, by its definition, is a monotonically decreasing function of $v_{min}$, one can in principle go to lower energies as well, but one may only place a lower bound on the predicted rate, rather than make a true prediction.} \emph{restricted to the energy range dictated by the appropriate velocity range} \emph{i.e.} $E\in [E^{(2)}_{low},E^{(2)}_{high}]$.
Analogous to the energy mapping above, we have a rate mapping,
\be
\frac{dR_1}{dE_1} \Longleftrightarrow g(v_{min}) \Longleftrightarrow \frac{dR_2}{dE_2},
\ee
with
\be
\frac{dR_2}{dE_R}\left(E_2 \right) = 
\frac{\kappa^{(2)}\mu_1^2}{\kappa^{(1)}\mu_2^2}
\frac{\sigma_2(E_2)}{\sigma_1\left(\frac{\mu_1^2\, M_T^{(2)}}{\mu_2^2 M_T^{(1)}} E_2\right)}
\frac{dR_1}{dE_R}\left(\frac{\mu_1^2\, M_T^{(2)}} {\mu_2^2\, M_T^{(1)}}E_2\right)~.
\label{eq:ratequal}
\ee
Equations (\ref{eq:emap}), (\ref{eq:geq}) and (\ref{eq:ratequal}) are the central results of this paper. They make no astrophysical assumptions, but only rely upon the assumption that an actual signal has been observed. 

We now focus on the SI case, since there are a greater number of experiments probing this scenario, but the analysis for SD is similar. In this (SI) case we can use (\ref{eq:xsecSI}) to rewrite (\ref{eq:ratequal}) in a simple form
\be
\frac{dR_2}{dE_R}\left(E_2 \right) = 
\frac{C_T^{(2)}}{C_T^{(1)}} \frac{F_2^2(E_2)}{F_1^2\left(\frac{\mu_1^2\, M_T^{(2)}}{\mu_2^2 M_T^{(1)}} E_2\right)} 
\frac{dR_1}{dE_R}\left(\frac{\mu_1^2\, M_T^{(2)}} {\mu_2^2\, M_T^{(1)}}E_2\right)~,
\label{eq:ratequalSI}
\ee
where we have introduced a target specific coefficient 
\be
C_T^{(i)}=\kappa^{(i)}\left(f_p\, Z^{(i)}+f_n\, (A^{(i)}-Z^{(i)})\right)^2~.
\label{eq:CTdefn}
\ee

\begin{figure*}
\begin{center}
\includegraphics[width=0.7 \textwidth]{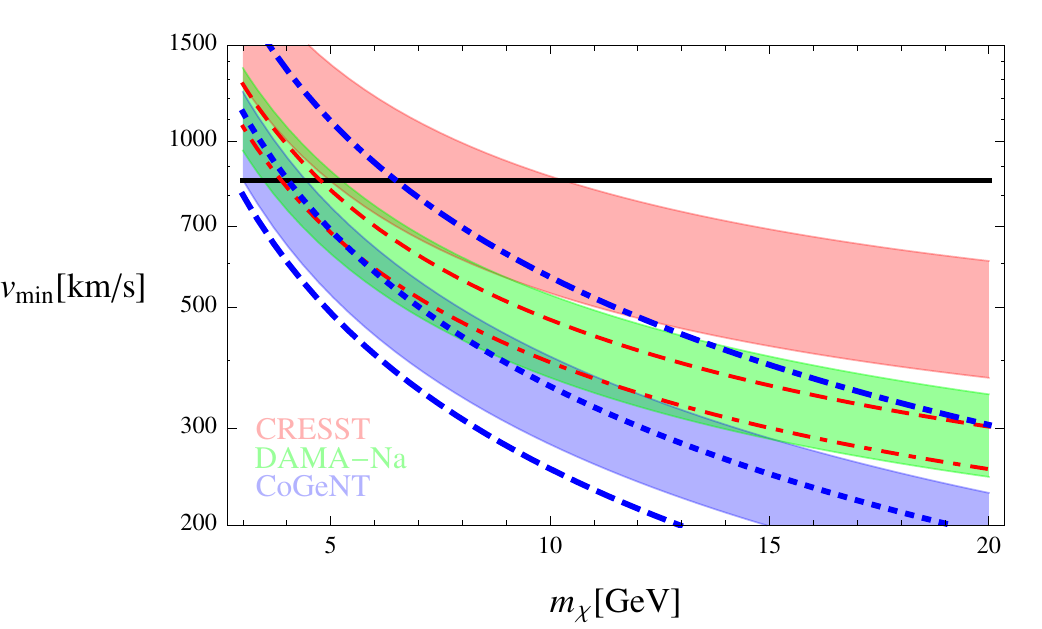}
\end{center}
\caption{
$v_{min}$ thresholds for various experiments. Solid bands are CRESST Oxygen band, 15-40 keV ({\em red, top}), DAMA Na band 6.7-13.3 keV ({\em green, middle}), CoGeNT Ge 1.9-3.9 keV ({\em blue, bottom}). Constraints are Xenon 1, 2 and 5 keV ({\em dashed, dotted, and dot-dashed, thick blue}), and CDMS-Si 7 and 10 keV, ({\em dot-dashed and dashed, thin red}).}
\label{fig:velocityplot}
\end{figure*}

In certain situations differential rates may not be available and instead it is only possible to compare total rates, this is the situation at present with CRESST.  In general the total rate at a particular experiment with energy --- and corresponding velocity --- thresholds of $(E_{low},v_{min}^{low})$ and $(E_{high},v_{min}^{high})$, can be expressed as,
\be
R = \frac{2 N_A \rho\, m_p}{m_\chi} \frac{\kappa}{M_T} \int_{v_{low}}^{v_{high}}dv\, \epsilon(E_R) \sigma(E_R(v)) v g(v)~.
\label{eq:integratedrate}
\ee
For the particular case of SI on which we are focused this becomes,
\be
R = \left(  \frac{2N_A \rho\, \sigma_p m_p}{m_\chi\,\mu_{n\chi}^2\, f_p^2} \right) \left(\frac{ \mu^2 C_T\, }{M_T} \right)  \int_{v_{low}}^{v_{high}}dv\, \epsilon(E_R) F^2(E_R(v)) v g(v)~,
\ee
where $\epsilon(E_R)$ an an energy-dependent efficiency. To compare two experiments, we must extract the energy dependent terms from the integral. So while we make no assumptions about $g(v)$, we evaluate the form factor at a value $\bar E_2= \bar E_1 \mu_2^2 M_T^{(1)}/\mu_1^2 M_T^{(2)}$ where the ratio $\epsilon_2(\bar E_2) F_2^2(\bar E_2)/\epsilon_1(\bar E_1) F_1^2(\bar E_1)$ is minimized or maximized, depending on whether we are considering a putative signal or constraint.  Thus comparisons of rates at two experiments may then be simply compared by taking ratios of $C_T$ with the form factor evaluated at the conservative value $\bar E$,
\be
R_2 \le \frac{\epsilon_2 (\bar E_2) F_2^2(\bar E_2)}{\epsilon_1 (\bar E_1) F_1^2(\bar E_1)} \frac{C^{(2)}_T}{C^{(1)}_T}\frac{M_T^{(1)}}{M_T^{(2)}}\frac{\mu^2_{2}}{\mu^2_{1}}R_1~.
\label{eq:CTeqn}
\ee
In order to determine what comparisons can be made between experiments, we must examine the relevant velocity space they probe. We re-emphasize that the signal at energy $E_{low}<E<E_{high}$ is sensitive to {\em all} particles with velocity greater than $v_{min}(E,M_N, M_\chi)$ through the integral $g(v_{min})$. A separate experiment with threshold $\tilde E$ will offer constraints independent of astrophysics if the resulting minimum velocity $\tilde v < v^{high}_{min,1}$. The optimal limits are reached when $\tilde v < v^{low}_{min,1}$. We illustrate this in Fig.~\ref{fig:velocityplot} for an ensemble of experiments, some with signals, some without.  The possible comparisons between these various experiments will be the subject of the subsequent sections.  Using (\ref{eq:ratequalSI}) scattering rates can be compared between experiments.  However, to compare to actual experimental data the relative exposures, efficiencies and other detector-specific factors must be correctly taken into account.  In the next section we describe in detail the experimental parameters necessary for the comparisons in the rest of the paper.

%%%%%%%%%%%%%%%%%%%%%%%%%%%%%%%
%%%%%%%%%%%%%%%%%%%%%%%%%%%%%%%%%%

\section{Applications: a comparison of existing experiments}
The important consequences of (\ref{eq:ratequal}) are immediately obvious. In principle, one can compare a positive signal at one experiment with one at another, or test the compatibility of a null result with a positive one. Unfortunately, ideal circumstances will rarely present themselves: additional backgrounds can complicate the extraction of $g(v)$, resolution can smear signals, or uncertainties in atomic physics (such as quenching factors) can complicate issues, making a precise extraction of the true $E_{NR}$ and hence $v_{min}$ impossible.  Furthermore, the signal may appear as a modulation (as in DAMA) limiting access to $g(v)$ to a summer/winter difference, or a lower bound on its mean.  Finally the signal may be of such low statistics that a reliable inference on the shape of $g(v)$ will be impossible, as is expected in many experiments before scaling to larger targets or running for longer exposures.

Nonetheless, in light of these challenging issues, there remain meaningful comparisons that can be made between experiments. Especially since these transformations preserve all information in the signal, we should be able to make the strongest possible relative statements without invoking additional assumptions about the halo. Such results are especially interesting in view of recent results that may pertain to light WIMPs. Since light WIMPs probe the highest part of the velocity distribution, where deviations from Maxwellian properties are the most likely, our approach is especially relevant.

We consider three potential signals: the CoGeNT low-energy excess \cite{Aalseth:2010vx}, the DAMA annual modulation signal \cite{Bernabei:2000qi} and the recently reported Oxygen-band events at CRESST \cite{idmtalkcresst} and for constraints: XENON10 (both conventional analyses and S2-only) and CDMS-Si; we describe the relevant parameters necessary for comparison between the various experiments below. 

\subsection*{CoGeNT}

The CoGeNT experiment \cite{Aalseth:2010vx} consists of a low noise germanium detector with 330 g of fiducial mass which has reported data for 56 days of exposure.  CoGeNT reports recoil energies that range from $\sim 0.4$ keVee to $\sim 12$ keVee, but we focus here on the events between 0.4 keVee and 3.2 keVee.  The observed electron equivalent energy is related to the nuclear recoil energy by, $E_{obs} = 0.2 (E_r/\kev)^{1.12}$, 
%\be
%E_{obs}= 0.19935\, E_r^{1.1204}~,
%\ee
so that the range of nuclear recoils of interest is $1.9-12\, \kev$.  In this range there are two cosmogenic peaks whose position and width are well understood, a relatively flat spectrum above these peaks and a clear excess at energies below the peaks.  It is this low energy excess that may be due to a DM signal and, rather than assume a particular functional form and fit, we extract it from the data by taking the data below the first peak ($E_{ee}<1$ keVee) subtracting from it the average of the high bins ($E_{ee}>1.6$ keVee).  Thus our ``signal" region, shown in Fig.~\ref{fig:velocityplot}, is $ 0.42\, \kevee\, (1.9\, \kev)< E_{ee} (E_r) < 0.92\,\kevee\, (3.9\,\kev)$.  Finally, we take into account the detector efficiency \cite{Aalseth:2010vx}.

\subsection*{DAMA}

DAMA, which has a NaI target, has accumulated a total of 1.17 ton-years of data, from both DAMA/NaI~\cite{Bernabei:2000qi} and DAMA/LIBRA \cite{Bernabei:2008yi,Bernabei:2010mq}.  They have observed, at the 8.9 $\sigma$ C.L., an annually modulating signal, in the 2-6 keVee energy range, whose phase is consistent with that expected from DM.  For low DM mass, recoils off the sodium dominate the spectrum, we take the quenching factor to be $q_{Na}=0.3$, although we will discuss the effects of varying this.  We concentrate here on the low energy range of recoil energies, 2-4 keVee, and in Fig.~\ref{fig:velocityplot} we show the $v_{min}$-space for this range of energies, assuming a quenching factor $q_{Na}=0.3$.  A lower (higher) $q_{Na}$ will result in the band moving higher (lower) in the plot. Specifically, for $q_{Na} \approx 0.45$ CoGeNT and DAMA are probing nearly identical ranges of velocity space.

In addition to the modulated rate DAMA has measured the total rate of recoil events and, as emphasized in \cite{Chang:2008xa}, the DAMA unmodulated rate may also provide non-trivial constraints on models of DM.  We do not consider the effects of channelling, which are believed to be small \cite{Bozorgnia:2010xy}.  We take into account the energy resolution at DAMA, by smearing with a Gaussian distribution of width $\sigma(E)/E = 0.448/\sqrt{E} + 0.0091$, with $E$ in keVee.

\subsection*{CRESST}
\label{sec:cresstvals}
CRESST consists of 9 CaWO$_4$ crystals (and 1 ZnWO$_4$ crystal).  They recently reported \cite{idmtalkcresst} an excess of O-band events in approximately 400 kg days of exposure.  Although they have not yet reported a detailed spectrum for each crystal it is still possible to gain some spectral information from Ref.~\cite{idmtalkcresst}.  

The upper bound on the energy of the events is set by their 40 keV upper limit on their search box. For the lower bound we use a threshold of 15 keV. In reality, the threshold in each of the nine detectors is determined by the value where the leakage is expected to be 0.1 events, which ranges from 9.65 to 22.65 keV.  In this total range 32 events are seen with an expected background of 8.7. In Ref.~\cite{idmtalkcresst} individual detectors are listed and two detectors are explicitly plotted,  and consequently we can determine that for seven detectors, there are 22 of 27 events above 15 keV, with an expected background (in these detectors) of 7.2 events, where to be conservative we have attributed all neutron and gamma backgrounds to these seven detectors.  Thus, since we only use data from 7 detectors, and taking into account an efficiency of 90\% (as used in the commissioning run), the data considered here has total exposure 280 kg days, and 22 events between 15-40 keV, with an expected background of 7.2.  In the future, when all thresholds, exposures, and events are reported by the CRESST collaboration, these results can be refined, but for the moment, we make these conservative assumptions.  The range of threshold velocities (i.e., $v_{min}$ values) corresponding to CRESST O-band events between 15-40 keV is shown in Fig.~\ref{fig:velocityplot}.

For constraints, we consider those experiments that are particularly sensitive to light WIMPs: XENON10 (both a conventional analysis \cite{Angle:2009xb} with a range of scintillation efficiencies, as we shall describe, as well as S2 only) and CDMS-Si.

\subsection*{XENON10}
We use the unblind XENON10 analysis \cite{Angle:2009xb} on a 5.4 kg active target of Xenon taken over 58.6 days between October 2006 and February 2007.  This analysis found 13 events between 16 keV and their upper threshold of 75 keV.  Their lower threshold is set by requiring a minimum of $\sim 12$ ionization electrons in the S2 signal, which for a constant $\mathcal{L}_{eff}=0.19$ corresponds to a threshold of $\sim 2\,\kev$, in addition there is an analysis threshold on the S1 signal of $\sim 5\ \kev$, with the same assumption on $\mathcal{L}_{eff}$.  We will also consider a potential S2-only low-threshold (7 drift electron $\sim$ 1 keV) experiment as recently discussed by \cite{sorensentalk}, we consider the efficiency adjusted exposure to be 5.1 kg days. For the charge yield considered in \cite{sorensentalk}, this would correspond to a threshold of 1 keV. In Figure~\ref{fig:velocityplot} we show limits for a Xenon experiment (thick blue lines) with a 5 keV threshold (upper line), as well as a what sensitivities 1 and 2 keV (lowest line and middle line, respective) thresholds would achieve. 

There is considerable uncertainty in the behavior of $\mathcal{L}_{eff}$, especially at low energies, and here we will consider the three cases discussed in \cite{Sorensen:2010hq} which we denote by MIN, MED and MAX ordered by increasing value of $\mathcal{L}_{eff}$ at recoil energies of 2 keV.  The corresponding detector resolutions and efficiencies are taken from \cite{Sorensen:2010hq}.

\subsection*{CDMS-Si}

The CDMS experiment contains both germanium and silicon detectors, we focus here on Si since it is sensitive to lighter WIMPs due to smaller mass and lower thresholds.  There have been several analyses of silicon data taken at the Soudan mine \cite{Akerib:2005zy,Akerib:2005kh,Filippini:2009zz} which combined have a raw exposure of 88.6 kg days.  For these analyses we use the efficiency presented in \cite{Filippini:2009zz} which has a threshold for nuclear recoils of 7 keV.  %In addition to the Soudan mine data, there was a smaller amount of data, 6.6 kg days of raw silicon data, taken at the Stanford Underground Facility \cite{Akerib:2003px} but with a recoil threshold of 5 keV.  
%In our analysis we use the efficiency function presented in Ref.~\cite{Savage:2008er}.  
In all data there are no signal events observed below 50 keV.  We do not consider here the recent low threshold analysis of CDMS-SUF~\cite{Akerib:2010pv}.

\vspace{0.2in}

In Fig.~\ref{fig:velocityplot} we show the regions of velocity space associated with the experiments discussed above, the bands denote the range of $v_{min}$ necessary to see events in the experiments with possible DM signals and the region between the curves are the same quantity at various null experiments.  We show both 7 and 10 keV thresholds for a silicon experiment, we choose these limits as at 7 keV the CDMS efficiency is going to zero, making limits that are reliant on that threshold more sensitive to uncertainties, while at 10 keV it is stable at $\sim$ 20\%.  For a Xenon target we consider 1, 2 and 5 keV thresholds, which may be reachable for different assumptions regarding $\mathcal{L}_{eff}$ and $Q_y$.  In Table~\ref{tab:ranges} we show, for fixed $m_\chi=10\, \gev$, how the energy ranges probed at CoGeNT, DAMA and CRESST translate to various elements used in other experiments.

A careful examination of Fig.~\ref{fig:velocityplot} shows a number of things: first, CoGeNT can be tested only by the 7 keV threshold of CDMS-Si, as well as a Xenon analysis sensitive to low energies. The first two have sensitivity below 10 keV, while Xenon can only make astrophysics-independent statements if the threshold is lower than 2.5 keV. This demonstrates, explicitly, that a model-independent comparison involves reaching signals present at 2 keV, and clarifying the scale to which $\mathcal{L}_{eff}$ must be measured. The CDMS-Si analysis is only sensitive right at its threshold to CoGeNT. As questions have been raised \cite{Hooper:2010uy} about the precise value of the Si threshold, if one restricts oneself to the higher threshold, no limits are possible.

Similarly, DAMA is tested well by CDMS-Si with a 7 keV threshold, but only marginally at 10 keV. Low-threshold Xenon analyses can give robust limits of DAMA, while higher thresholds are generally limited to heavier masses. 

Finally, we see that the CRESST results are completely tested by the low-threshold XENON10 analysis, CDMS-Si (even with a 10 keV) threshold. While the nominal threshold, depending on the details of $\mathcal{L}_{eff}$, of XENON10 ($\sim$ 5 keV) and XENON100 ($\sim$ 6 keV) is too high, both experiments can probe down to 4 keV with moderately reduced sensitivity, and energy smearing will given XENON sensitivity to the CRESST signal.

With these ranges in hand, we can proceed to compare the experiments directly. We shall see that if the potential signal is large enough, $g(v)$ can be extracted directly, even if $f(v)$ cannot be extracted with any reliability. In such cases, we can make slightly stronger statements involving the spectra. However, even if $g(v)$ cannot be reconstructed, we can still make significant statements by integrating over the relevant velocity range.

\begin{table}[t]
   \centering
   \begin{tabular}{c|cccccc} 
    Approx. range & O & Na & Si & Ar & Ge & Xe \\
   \hline
   CoGeNT (Ge): 2 - 4   & 4.3 - 8.6 & 3.9 - 7.8 & 3.6 - 7.2 &  3.0 -  6.0 & 2 - 4       & 1.3 - 2.5  \\
   DAMA (Na): 6 - 13     & 6.6 - 14  & 6 - 13     & 5.5 - 12  & 4.6 - 10    & 3.1 - 6.7  & 1.9 - 4.2 \\
   CRESST (O): 15 - 40 &15 - 40   & 14 - 36    & 12 - 33  & 10 - 28     & 6.9 - 19   & 4.3 - 12
   \end{tabular}
   \caption{Conversion of energy ranges (all in keV) between various experiments/targets for a 10 GeV DM particle, using the expression in (\ref{eq:emap}).}
   \label{tab:ranges}
\end{table}

\subsection{Application I: Employing Spectra in Near-Ideal Situations (CoGeNT)}

We consider first the situation when there is sufficient data to be able to extract a recoil spectrum,
CoGeNT is a example of such an experiment, because the putative signal is quite large.   We concentrate on the events below 3.2 keVee where the DM signal should be largest and there are few cosmogenic backgrounds.  In this range, in addition to the possible DM signal at low energies, the data contains several clear cosmogenic peaks and a constant background above the peaks.  We average the [1.62-3.16 keVee] bins as an estimate of the constant background and subtract this from the bins in the [0.42-0.92 keVee] range, which we then consider as the DM signal, after this subtraction there are 92 signal events before efficiency correction. This allows us to determine $g(v)$ or, equivalently, predict the rate at any other experiment in the equivalent energy range. One can easily observe from its definition that $g(v)$ is monotonically decreasing as a function of $v$ (see, for instance the discussion in~\cite{Fox:2010bu}), and thus the value at the low end of this range is a lower bound for lower values of $v$. This is not especially relevant for our analysis here, but would be likely relevant in situations where the other experiments could probe lower energies as well.

%%%%%%%%%%%%%%%%%%%%%%%%%%%%%%%%

\begin{figure}[t] %  figure placement: here, top, bottom, or page
   \centering
   \includegraphics[width=0.85\textwidth]{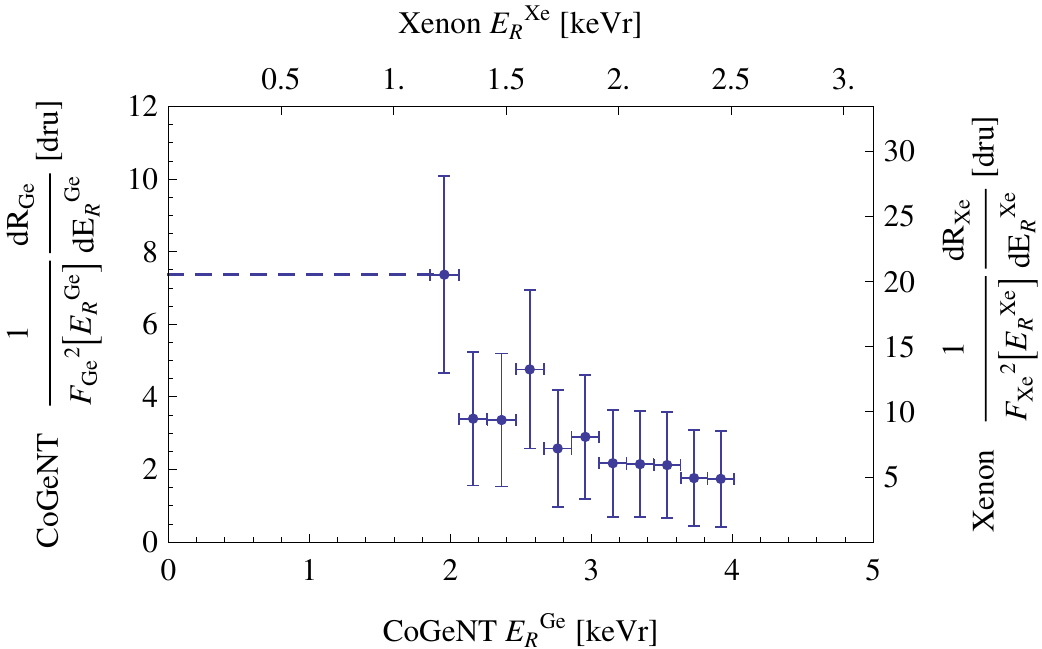} 
   \caption{The extracted CoGeNT signal (left and bottom axes) and the rate it is mapped to on a Xenon target (top and right axes) for $m_\chi = 10\, \gev$ (rescaled by form factors at the corresponding energies $F^2_{Xe}(E_R^{Xe}),F^2_{Ge}(E_R^{Ge}) \sim 1$). The dashed line is the lower bound on the rate at low energies, using the monotonically falling nature of $g(v_{min})$.}
   \label{fig:mappedcogent}
\end{figure}
%%%%%%%%%%%%%%%%%%%%%%%%%%%%%%%%

Since we will compare this with the XENON10 experiment, we choose $f_p=1$ and $f_n=0$, which is motivated from light mediators mixing with the photon, since it will give the most lenient bounds.  Using (\ref{eq:ratequalSI}) we can map the CoGeNT signal onto a Xenon target, and study the signal that would arise at XENON10.  We show this in figure \ref{fig:mappedcogent}.

What is remarkable about this figure is that -- once the CoGeNT signal is specified -- the expected rate on a Xenon target is completely unambiguous (and similarly on any other target). This involves no assumptions about the halo escape velocity, velocity dispersion, or even the assumption that the velocity distribution is Maxwellian, but requires only an input of the WIMP mass.

\begin{figure}[t] %  figure placement: here, top, bottom, or page
   \centering
   \includegraphics{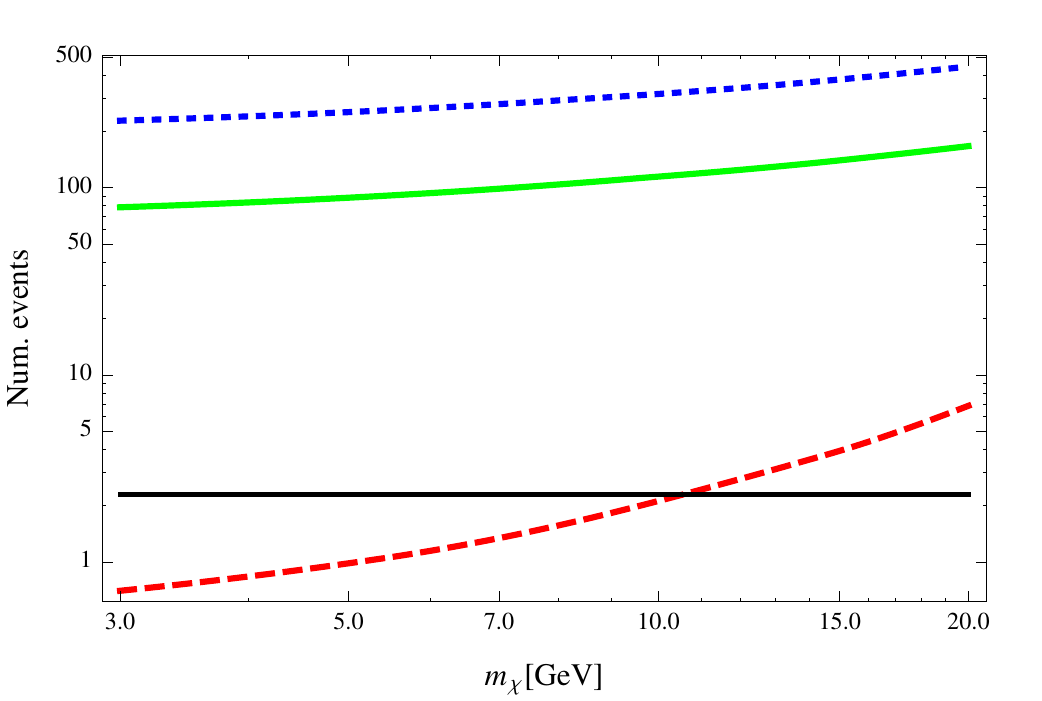} 
   \caption{The number of events predicted at XENON10 by the possible DM signal at CoGeNT for 3 cases of $\mathcal{L}_{eff}$, MIN (dashed red), MED (solid green) and MAX (dotted blue).  The black line is the 90\% C.L. upper limit on the number of events allowed by XENON10 data, the region above this line is excluded at 90\% confidence.}
   \label{fig:cogenttoxenon}
\end{figure}

After taking into account exposure and the detector efficiencies (MIN, MED and MAX cases described above) we can predict the total number of events predicted by the CoGeNT events (if they are indeed coming from elastically scattering DM), we show this in Fig.~\ref{fig:cogenttoxenon}.  Since there were no events at XENON10 in the energy range corresponding to the CoGeNT range we see that \emph{independent of all astrophysical assumptions}, only for $L^{MIN}_{eff}$ are CoGeNT and XENON10 are consistent at the 90\% C.L.  In the MIN case, $m_\chi < 11\ \gev$ allows CoGeNT to evade XENON10. For MED and MAX cases the predicted signal at XENON10 would be too large by a significant amount, excluding the elastic SI WIMP scattering interpretation by more than an order of magnitude. 
%%%%%%%%%%%%%%%%%%%%%%%%%%%%%%%%

%%%%%%%%%%%%%%%%%%%%%%%%%%%%%%%%

Because of the uncertainties associated with extraction of the value of $\mathcal{L}_{eff}$ at low energies, additional attempts have been made to probe the low energy region with Xenon experiments. In particular, \cite{sorensentalk,inprogresspeter} examined data from XENON10, and used only the ionization signal (S2), which is typically larger than S1 and can allow a more reliable signal at low energies. The value of the charge yield (drift electrons per keV) was extracted from Monte Carlo. Using the values there, the equivalent energy range for CoGeNT is approximately $8\sim13$ electrons, above the 7 electron threshold. Assuming a value of $Q_y = 4\ {\rm electrons/keV}$ for instance, the threshold of 7 electrons at XENON10 only captures a portion of the signal predicted by CoGeNT. 

\begin{figure}[t] %  figure placement: here, top, bottom, or page
  \includegraphics[width=0.45\textwidth]{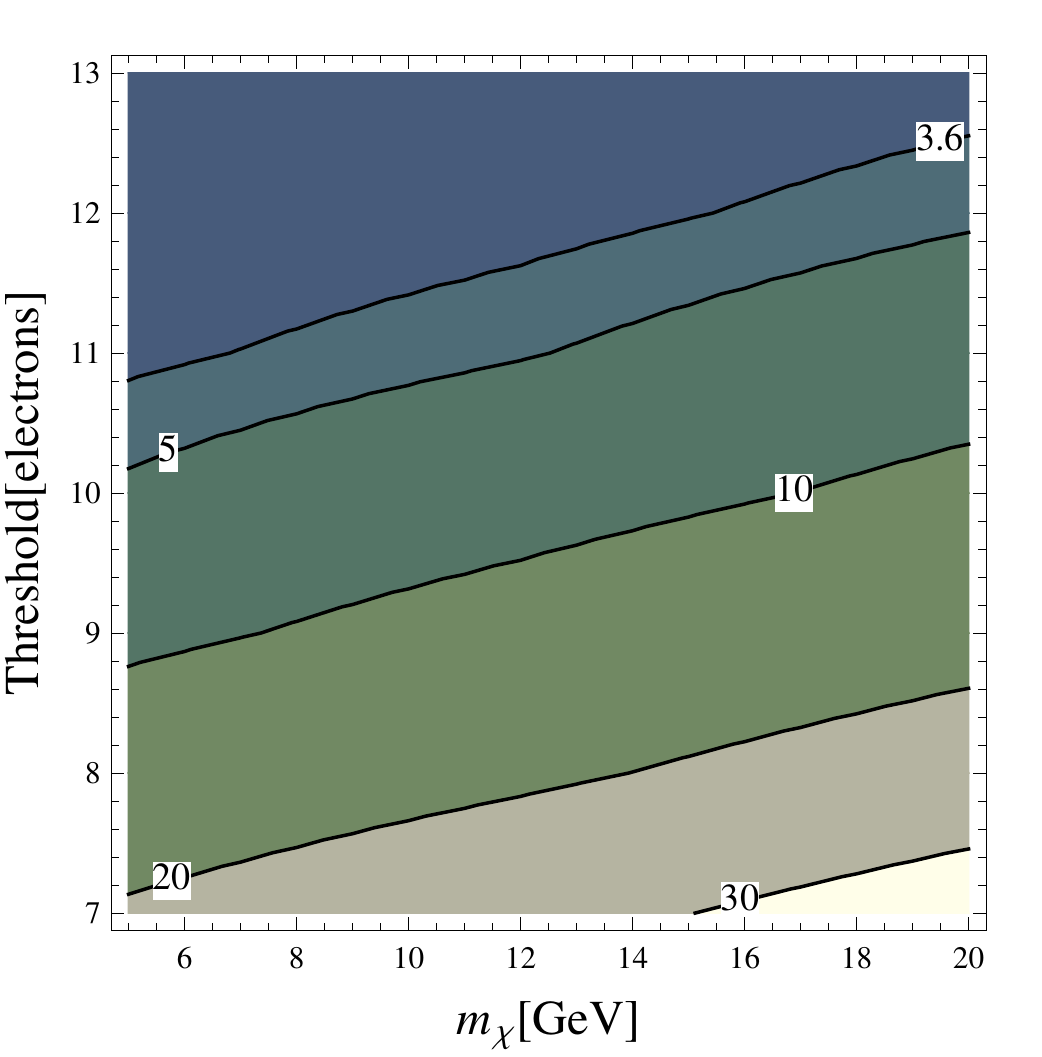} 
  \includegraphics[width=0.45\textwidth]{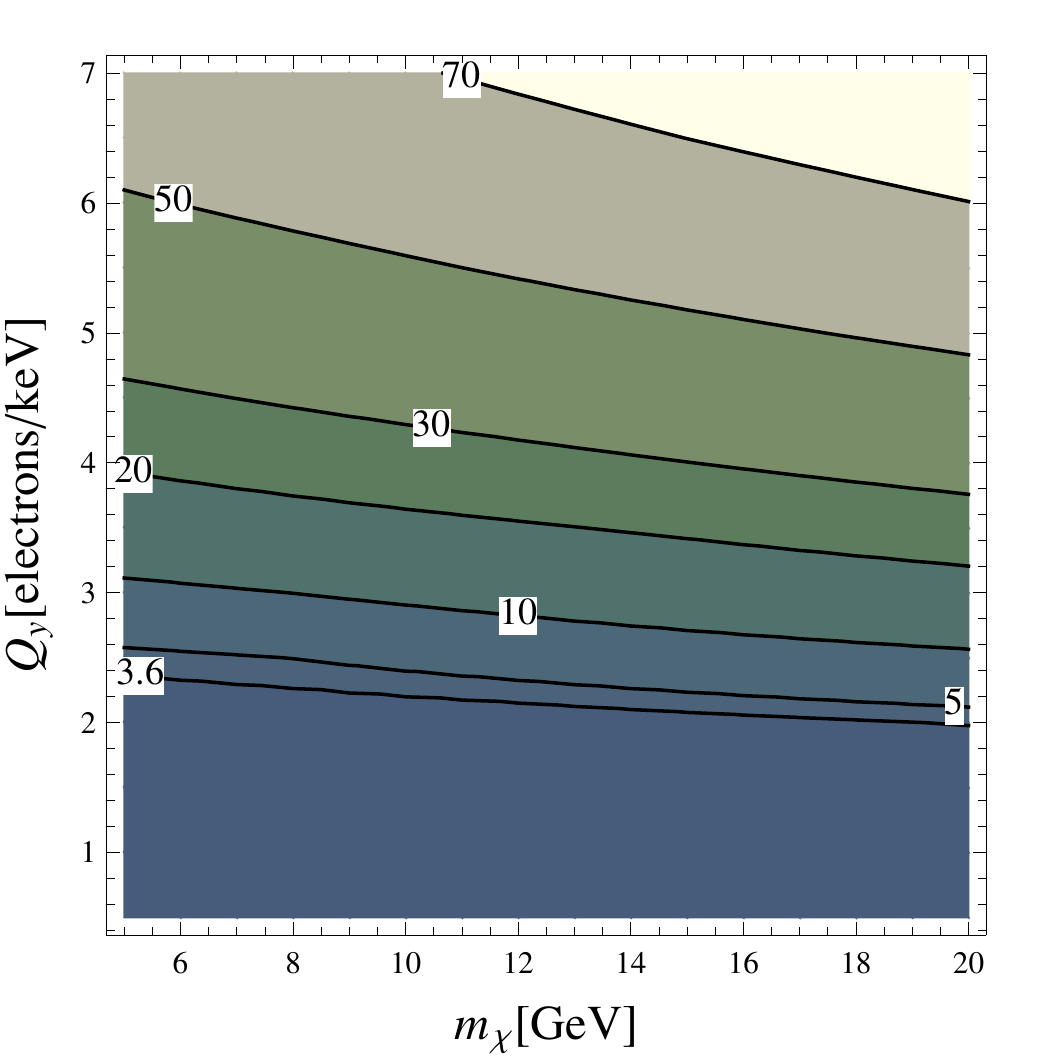} 
   \caption{({\em left}) The number of events predicted (labels on contours), by CoGeNT, at XENON10 for an S2 only analysis \cite{sorensentalk,inprogresspeter} for various S2 thresholds, assuming a constant value $Q_y = 4 {\, \rm electrons/keV}$. ({\em right}) The signal above threshold of 7 electrons, but assuming different constant values of charge yield, $Q_y$.}
   \label{fig:cogenttoxenons2}
\end{figure}

While the 7 electron cutoff corresponds to a particular value of energy in principle, Poisson fluctuations smear this. Nonetheless, an interesting question is the expected rate on the target used by \cite{sorensentalk,inprogresspeter}, with 5.1 kg d of effective exposure. %We show in Figure \ref{fig:cogenttoxenons2} the expected rate as a function of this low energy cutoff. One sees that only if the energy range is entirely below what should be seen at XENON is it possible for the CoGeNT signal to be consistent.
This is most easily phrased in terms of the question of what charge yield can make these experiments consistent. Assuming a constant charge yield over the energies in question, we can calculate the likelihood based on Poisson fluctuations of events appearing in the XENON10 experiment, which we show in Figure~\ref{fig:cogenttoxenons2}. One sees that one would require a charge yield of roughly $Q_y \lsim 2.4\, {\rm electrons/keV}$ for consistency, much lower than the value of $Q_y \approx 7$ extracted by \cite{sorensentalk,inprogresspeter}. Whether such a significant difference is reasonable will no doubt be subject to a great deal of discussion \cite{Collar:2010ht}.

%%%%%%%%%%%%%%%%%%%%%%%%%%%%%%%%

%%%%%%%%%%%%%%%%%%%%%%%%%%%%%%%%

\subsection{Application II: Total Rate Comparisons in Sub-Optimal Situations (CRESST)}
\begin{figure}[t] %  figure placement: here, top, bottom, or page
   \centering
   \includegraphics[width=0.45\textwidth]{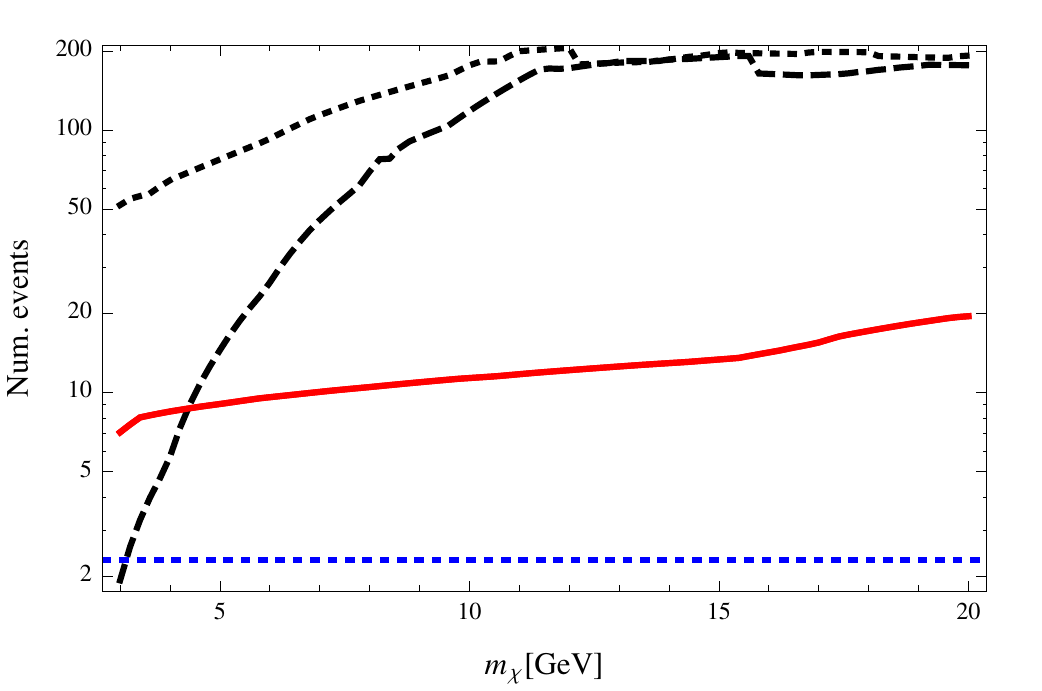} 
   \includegraphics[width=0.45\textwidth]{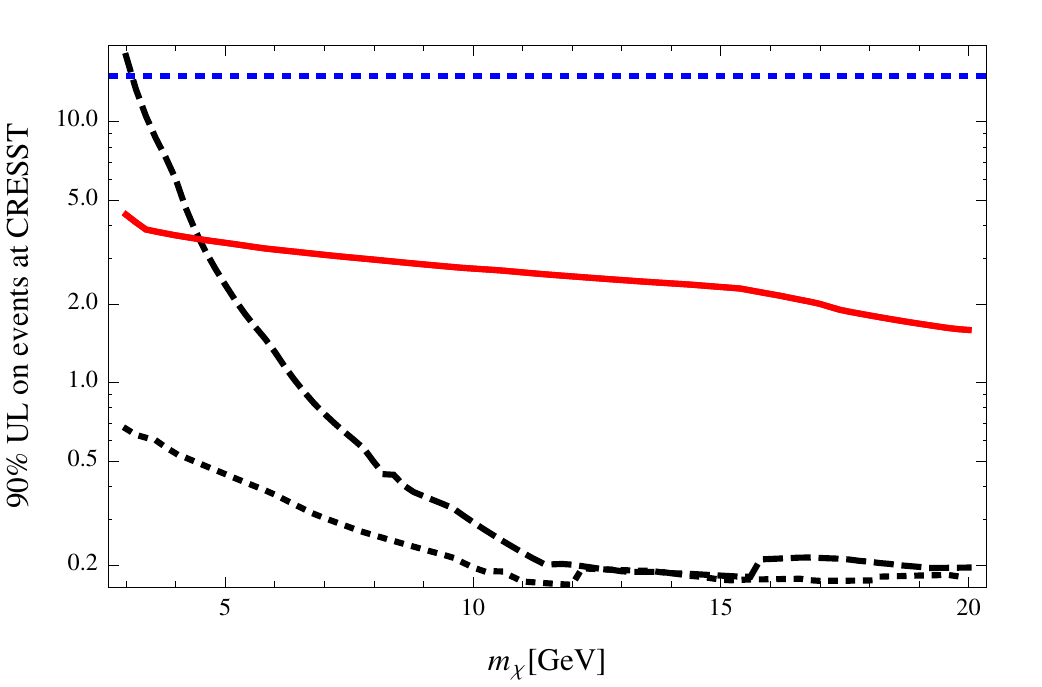} 
   \caption{LH plot: the CRESST prediction for the total number of events at CDMS-Si (solid red) and XENON10, for $\mathcal{L}_{eff}^{MIN}$ (dashed black) and $\mathcal{L}_{eff}^{MED}$ (dotted black), the dotted (horizontal blue) line is the 90\% C.L. upper limit on the number of events allowed by CDMS-Si, the region above this line is excluded at 90\% confidence.  RH plot: the 90\% C.L. upper limit on the number of events at CRESST as predicted by CDMS-Si (solid red) and XENON10, again for $\mathcal{L}_{eff}^{MIN}$ (dashed black) and $\mathcal{L}_{eff}^{MED}$ (dotted black), the dotted (blue) line is the number of events we estimate above background in CRESST.}
   \label{fig:cresstcomparisons}
\end{figure}

The above situation with CoGeNT is close to ideal: low backgrounds, high statistics, good energy resolution and calibration. In contrast, there are often situations with significantly less ideal characteristics. In particular, it may be that not enough is known about the backgrounds, or the data itself, to be able to extract a recoil spectrum for DM, but we shall see it is nonetheless possible to say something about the total number of DM scatters.  This is the case for the CRESST data, which we estimate has 15 events above background between 15 and 40 keV (see the discussion in \ref{sec:cresstvals}).  We use (\ref{eq:CTeqn}) to compare the CRESST integrated rate to the null results of both CDMS-Si and XENON10, Fig.~\ref{fig:cresstcomparisons}.  When comparing the two experiments we take into account efficiencies and form factors so as to be as conservative as possible, as explained after (\ref{eq:CTeqn}).  

%%%%%%%%%%%%%%%%%%%%%%%%%%%%%%%%%
%%%%%%%%%%%%%%%%%%%%%%%%%%%%%%%%%%%%%

As is clear from Fig.~\ref{fig:cresstcomparisons} any sizeable signal in this range is highly incompatible with both the XENON10 and CDMS-Si results. While some have criticized the calibration at the lowest energies for CDMS-Si \cite{Hooper:2010uy}, the lowest energy relevant for 15 keV Oxygen recoils is above 10 and typically 11 keV on Silicon, depending on the WIMP mass. Thus, these constraints are likely quite stable to future modifications, making elastic WIMP scattering very unlikely to be the explanation of the CRESST anomalous events.

\section{Other Applications and Future Results}

\begin{figure}[t] %  figure placement: here, top, bottom, or page
   \centering
   \includegraphics[width=0.45\textwidth]{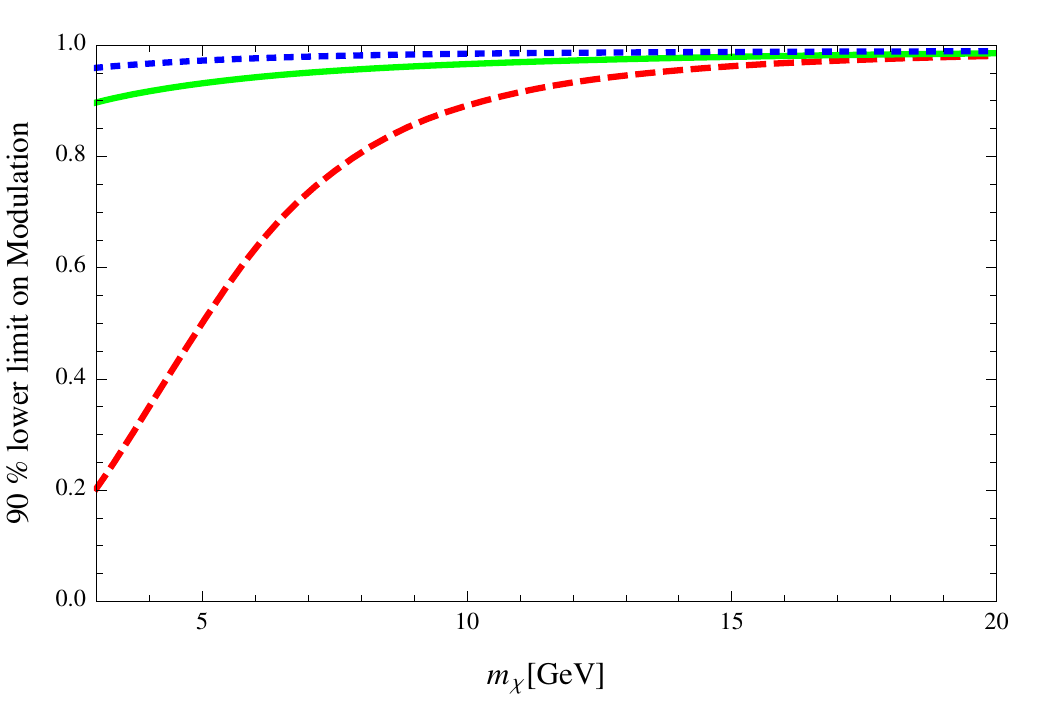} 
   \includegraphics[width=0.45\textwidth]{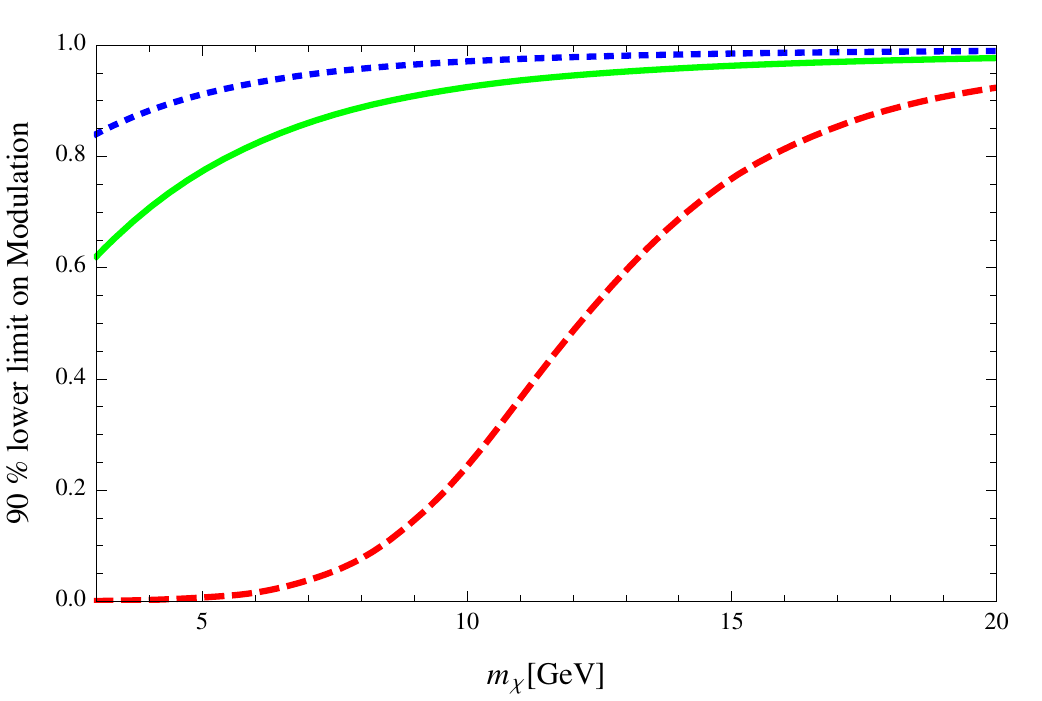} 
   \caption{The 90\% C.L. lower limit on the modulation fraction allowed by XENON10 data, for a quench factor in sodium of 0.3 (LH plot) and 0.45 (RH plot) and for 3 cases of $\mathcal{L}_{eff}$, MIN (dashed red), MED (solid green) and MAX (dotted blue).  A modulation fraction below the curves is ruled out at 90\% confidence.}
   \label{fig:damatoxenon}
\end{figure}

DAMA also has extracted a recoil spectrum, possibly associated with DM, but in this case it is for the modulating part of the DM signal, i.e. DAMA allows extraction of $g(v,t)$.  We can repeat the exercise of translating from one experiment to another to get a prediction for the size of the modulating signal at XENON10.  Since XENON10 took its data in the winter and saw no events in the region corresponding to DAMA's 2-6 keVee, this places an upper limit of 2.3 events in the winter which in turn places a lower bound on the amount of modulation the DM signal must have in order not be ruled out by XENON10's null result.  We present this lower bound on the modulation fraction\footnote{We define the modulation fraction as $\frac{S-W}{S+W}$ where $S,W$ denote the summer and winter event rate respectively.} in Fig.~\ref{fig:damatoxenon} for two choices of the quench factor in sodium, $q_{Na}=0.3, 0.45$.  Thus, irrespective of astrophysics, in order for DAMA to be consistent with XENON10 the modulation fraction has to be larger than 20\% and in most cases almost 100\% for the standard assumption of $q_{Na}=0.3$.  For the more extreme choice of $q_{Na}=0.45$ the modulation may be smaller but for DM heavier than 10 GeV it again has to be above 20\%.

%%%%%%%%%%%%%%%%%%%%%%%%%%%%%%%%%
%%%%%%%%%%%%%%%%%%%%%%%%%%%%%%%%%%%%%

An interesting relationship between CoGeNT and DAMA can be made here. The modulation at DAMA can be applied to CoGeNT through (\ref{eq:CTeqn}). In doing so, one finds a modulation $O(0.8 - 0.9 {\rm cpd/kg})$ expected at CoGeNT. With a quenching factor $q_{Na} = 0.3$, this is expected to overlap the L-shell peaks, which, in decaying away, would make a rising signal difficult to extract. The modulation in the signal range we cannot predict.

On the other hand, if $q_{Na} = 0.45$, then the energy range of CoGeNT overlaps that of DAMA. The $0.8 - 0.9 {\rm cpd/kg}$ modulation amplitude would then be visible over the $\sim 5 {\rm cpd/kg}$ in that range (i.e., a modulated amplitude of $\sim 20$ \%, or $\sim 40$\% peak-to-peak), which should be detectable over an annual cycle.

Finally, it is intriguing to employ this technique to study what sensitivities future experiments will have to existing signals. In particular, we can consider the COUPP experiment, with a ${\rm CF_3I}$ target. Focusing on scattering off of Fluorine, the CoGeNT events would be visible in a range 4.7 -- 9.4 keVr, while the DAMA modulation would be present in the range 7.3 -- 14.6 (4.8 -- 9.7) keVr for $q_{Na} = 0.3 (0.45)$, for $m_\chi = 7$ GeV. Thus, a threshold in the 5 -- 9 keVr range should allow astrophysics independent tests of these signals. 

\section{Discussion}
The search for WIMP dark matter has as its elements three central goals: to discover the WIMP, and to measure its mass and interaction cross section with matter. Although we have proceeded for years without a confirmed discovery, the focus has been on what ranges of mass and cross section are excluded. Regrettably, this thinking has crept into our whole approach to discussing a comparison of WIMP searches --- we compare compatibility within the confines of $m_\chi - \sigma$ plots, confusing the answer to the latter two questions (the WIMP mass and properties) with the central first question: has dark matter been discovered? To do so necessarily entangles our astrophysical uncertainties into our results, and even worse, makes it difficult, if not impossible, to determine how sensitive the conclusions are to variations in the halo model.

To this end, we have explored a new technique to compare experiments in the presence of a positive signal, and importantly, to do so without invoking {\em any} astrophysical model whatsoever. The energy range of a given experiment can, for a given WIMP mass, be mapped unambiguously into an equivalent energy range at another using the expression in (\ref{eq:emap}), allowing an apples-to-apples comparison of signals and limits. Moreover, the measured rate at one experiment can be mapped into a rate at the other experiment with the expression in (\ref{eq:ratequal}), once the particle physics model is specified, yielding a completely unambiguous prediction for the second experiment. 

\begin{figure}[t] %  figure placement: here, top, bottom, or page
   \centering
   \includegraphics[width=0.95\textwidth]{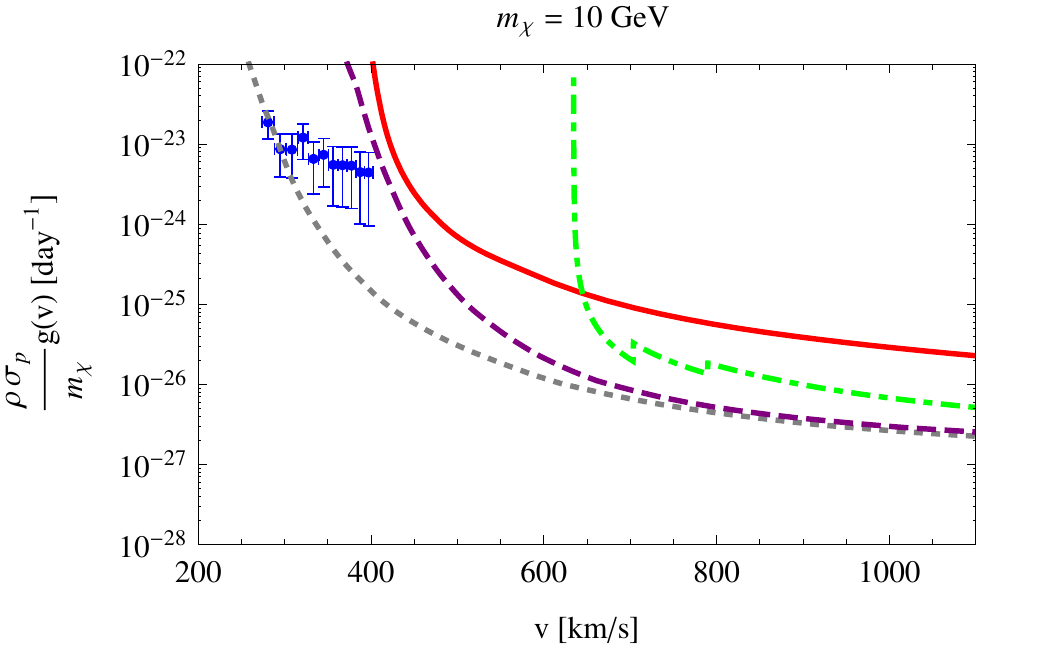} 
   \caption{A comparison of measurements and constraints of the astrophysical observable $g(v)$ [see relevant expressions in (\ref{eq:rate}),(\ref{eq:defgofv}),(\ref{eq:geq})] for $m_\chi = 10\ \gev$: CoGeNT (blue), CDMS-Si (red, solid), CDMS-Ge (green, dot-dashed), XENON10 - MIN $\mathcal{L}_{eff}$ (purple, dashed), and XENON10 - MED $\mathcal{L}_{eff}$ (gray, dotted). CoGeNT values assume the events arise from elastically scattering dark matter, while for other experiments, regions above and to the right of the lines are excluded at 90\% confidence. The jagged features  of the CDMS-Ge curve arise from the presence of the two detected events.}
   \label{fig:gofv}
\end{figure}

This is done by implicitly solving for the function $g(v)$. In a sense, each experiment is actually a measurement or upper limit of the function $g(v)$. This motivates a new and simple comparison of experimental results by simply showing the different values and limits extracted for this function from different results, which we do in Fig.~\ref{fig:gofv}. 
%%%%%%%%%%%%%%%%%%%%%%%%%%%%%%%%%
%%%%%%%%%%%%%%%%%%%%%%%%%%%%%%%%%%%%%

To determine this plot, in the presence of a positive signal, one needs merely to read off $g(v)$ from (\ref{eq:geq}). In the absence of a (clear) signal, there is always a certain element of choice in how one quantifies a constraint. However, one can exploit the fact that $g$ is a monotonically decreasing function, so for our constraints, we simply assume that $g(v)$ is constant below $v$, and assume a Poisson limit on the integral of (\ref{eq:geq}) from the experimental threshold to $v$. However, other techniques could also be used, see the Appendix for more details.
 
This approach with a $g-v$ plot has numerous advantages over the traditional $m_\chi-\sigma$ plots. It makes manifest what the relationships between the different experiments are in terms of what $v_{min}$-space is probed, and shows (for a given mass) whether tensions exist. Moreover, the quantity $g(v)$ is extremely tightly linked to the data, with only a rescaling by form factor as in (\ref{eq:geq}). Thus, unlike $m_\chi-\sigma$ plots, which have a tremendous amount of processing in them, this provides a direct comparison of experimental results on the same plot.

In light of this, we would propose that all future results take a three-tiered approach to data comparison and presentation
\begin{itemize}
\item{{\em $m_\chi - \sigma$ plots:} the current standard presentation is extremely useful for understanding implications for models irrespective of other experiments, and for making model-dependent comparisons of different results. We emphasize that our new techniques are {\em complementary} to this, and not a replacement for it.}
\item{{\em A mapping of contested or compared results:} by employing (\ref{eq:ratequal}) an experiment can state in a completely astrophysics-independent fashion whether two results are compatible or not. The use of these tools would remove any discussion as to whether different assumptions about escape velocities would impact the question of consistency of conclusions.}
\item{{\em A measurement of $g(v)$:} By presenting results in a $g-v$ plot as in Fig.~\ref{fig:gofv}, the astrophysical dependences of results become more obvious, and relative (in)consistencies clear. For experiments with only a single nuclear target, the plot can be made for a single value of $m_\chi$, and can be unambiguously mapped to any other value. For experiments with multiple targets, this will be slightly more involved, but still will likely require plotting the result for only a few values of $m_\chi$.}
\end{itemize}

In the asymptotic future, we certainly hope for a large number of experiments, with different targets and technologies, all seeing results with high statistics. In the meantime, the question of whether dark matter has been discovered prompts the need for new techniques that do not rely on outside assumptions, but at the same time, keep the maximal amount of information possible. We have proposed such a technique, by mapping signals from one experiment to another via (\ref{eq:ratequal}), comparable energy ranges through (\ref{eq:emap}), and by extracting the {\em physical} astrophysical quantity $g(v)$ through (\ref{eq:geq}). 

We have shown numerous applications of this to existing results from CoGeNT, DAMA and CRESST. In particular, we have found that only for low values of $\mathcal{L}_{eff}$, a positive signal at CoGeNT can coexist with null results from the conventional XENON10 analyses. For S2-only analyses, however, we have seen that consistency requires a charge yield of $Q_y < 2.4\ {\rm electrons/keV}$, well below most current assumptions. By comparing DAMA and XENON10, for $q_{Na} = 0.3$, we have shown the modulation rate is typically higher than $80\%$, which would be difficult to achieve for elastically scattering WIMPs over a large range of $2-4$ keVee, even with non-standard halo models. On the other hand, for $q_{Na} = 0.45$, the modulation can be lower, $\sim 10\%$. However, in this case, the DAMA modulation maps into the CoGeNT energy range, predicting a clear modulation amplitude of $\sim 20 \%$ in the low-energy range. Finally, we have seen how even in situations with large backgrounds and unclear spectra, as is the case of the excess events reported by the CRESST collaboration, constraints by CDMS-Si and XENON10 unambiguously test this, making the elastic WIMP scattering explanation of them very unlikely.

While it remains to be seen which if any of these signals will turn out to be genuine signals of dark matter, these tools will provide a means to remove one of the important uncertainties in their comparison. Perhaps, with higher statistics, this can be inverted, and by requiring the consistency from different measurements of $g(v)$, dark matter will yield information on astrophysics, as well.

\section*{Acknowledgements}  We thank Juan Collar, Rafael Lang, David Moore and Peter Sorensen for useful comments on the manuscript. NW is supported by NSF grant PHY-0449818 and DOE OJI grant \# DE-FG02-06ER41417, as well as by the Amborse Monell Foundation.  Fermilab is operated by Fermi Research Alliance, LLC, under Contract DE-AC02-07CH11359 with the United States
Department of Energy. PF and NW would like to thank Aspen Center for Physics where this work was initiated.

\appendix
\section{Displaying direct detection results in an astrophysics independent fashion}

As well as allowing comparison of positive results between different experiments in a fashion that is independent of astrophysics, our technique also allows constraints to be compared to each other and to putative discoveries. Such comparisons can be made, such as we have done by constraining $g(v)$ in Figure~\ref{fig:gofv}.  Here we outline, in more detail, how this is carried out.

For positive results the comparison can be made at the spectrum level through the application of (\ref{eq:ratequal}). This is most easily done in the situation that the statistics are large enough, and backgrounds low enough, that a meaningful rate $dR/dE_R$ can be extracted. Then, using (\ref{eq:ratequal}), a direct measurement of $g(v)$ can be given. In situations where rates are too low to simply read off $g(v)$, alternative techniques would be needed. The simplest would just be to take a large enough bin in $v$ such that statistics are adequate, but more sophisticated approaches, utilizing the monotonicity of $g(v)$ would also be possible. We leave such studies for future work.

If an experiment does not see a sufficient number of signal events to claim discovery, then it is likely that one will wish instead to place a constraint on the properties of dark matter. In general, one should first ascertain the bound on the parameterized WIMP cross section at the confidence level required. This can be simply done, using whatever confidence estimator is already used for astrophysics-dependent $\sigma-m_\chi$ plots.

Suppose that one wishes to employ some confidence estimator $C(dR/dE_R(m_\chi))$ to place limits, where $ dR/dE_R(m_\chi)$ is the expected recoil spectrum for some $\sigma_0$, i.e., (\ref{eq:rate}). This estimator may simply be using Poisson statistics, evaluating the integral of the spectrum, or using more advanced techniques that use spectral information as well, such as those of Yellin~\cite{Yellin:2002xd}. For a given value of $m_\chi$, for instance, one varies other parameters until one achieves, e.g., $C=0.1$ allowing one to claim a 90\% exclusion for those parameters.  Assuming that such an analysis has already been performed for explicit halo models, it is straightforward to place a bound on $\rho \sigma g(v_{min})/m_\chi$, for a particular choice of DM mass, in the general astrophysics case.  

For standard $\sigma-m_\chi$ plots, $g(v)$ is fixed, for instance a Maxwellian distribution, with a fixed $v_0$ and $v_{esc}$. The only free parameters in  $ dR/dE_R(m_\chi)$ are then $m_\chi$ and $\sigma_0$ (in the SI case), or $a_p$ and $a_n$ (in the SD case). In our case, since we do not want to use a Maxwellian $g(v)$, we have an additional free parameter. 

Since $g(v)$ is a monotonically decreasing function an upper bound on its value at some velocity $v_1$, $g(v_1)\le g_1$, also applies to all lower velocities.  Thus, the most conservative form that the upper bound on $g(v)$ can take is that of a step function
\be
g(v;v_1)=g_1 \Theta(v_1-v)~.
\label{eq:stepfunctionform}
\ee  
Physically, this would correspond to stream in $f(v)$ with velocity $v_1$. 

Using this, (\ref{eq:rate}) becomes
\be
\frac{dR}{dE_R} = \frac{N_T M_T \rho}{2 m_\chi\mu^2}  \sigma(E_R)\, g_1 \Theta(v_1-v_{min}(E_R))~.
\label{eq:newrate}
\ee

For a given WIMP mass $m_\chi$, the overall scaling is now proportional to e.g.,  $\rho \sigma g_1/m_\chi$ in the SI case, rather than simply $\rho \sigma/m_\chi$ as in the standard case where $g$ is specified. For a given $v_1$, one can then place a limit on this combination using the existing estimator.

In short: to calculate the appropriate limits on $g(v)$, one should use whatever technique one was intending to use for the standard analysis, but now replace the Maxwellian $g(v)$ with the step function form. For any given $m_\chi$, one places a limit on  $\rho \sigma g_1/m_\chi$ as one would have on $\rho \sigma /m_\chi$, or, $\sigma$ for fixed $\rho$ and $m_\chi$, precisely as before.

%%%%%%%%%%%%%%%%%%%%%%%%
\bibliography{astroindependent}

\end{document}